\documentclass[preprint,5p]{elsarticle}
\usepackage{wrapfig}
\usepackage{mathtools, cuted}
\usepackage{amsmath}
\usepackage{lineno,hyperref}
\usepackage{xcolor}
\modulolinenumbers[5]

\journal{Ultramicroscopy}

\biboptions{square,comma,numbers,sort&compress}
\begin{document}

\begin{frontmatter}

\title{Plasmon energy losses in shear bands of metallic glass}

\author[IMP]{Maximilian Grove}
\author[IMP]{Martin Peterlechner}
\author[IMP]{Harald R\"osner\corref{correspondingauthor}}
\ead{rosner@uni-muenster.de}
\author[FEI]{Robert Imlau}
\author[ITA,CAM,CAV]{Alessio Zaccone}
\author[IMP]{Gerhard Wilde}

\cortext[correspondingauthor]{Corresponding author}

\address[IMP]{Institut f\"ur Materialphysik, Westf\"alische Wilhelms-Universit\"at M\"unster, Wilhelm-Klemm-Str. 10, 48149 M\"unster, Germany}
\address[FEI]{Thermo Fisher Scientific, Achtseweg Noord 5, 5651 GG Eindhoven, The Netherlands}
\address[ITA]{Department of Physics ''A. Pontremoli``, University of Milan, via Celoria 16, 20133 Milano, Italy}
\address[CAM]{Department of Chemical Engineering and Biotechnology, University of Cambridge, Philippa Fawcett Drive, CB3 0AS Cambridge, U.K.}
\address[CAV]{Cavendish Laboratory, University of Cambridge, JJ Thomson Avenue, CB3 9HE Cambridge, U.K.}

\begin{abstract}
Shear bands resulting from plastic deformation in cold-rolled Al$_{88}$Y$_{7}$Fe$_{5}$ metallic glass were observed to display alternating density changes along their propagation direction. Electron-energy loss spectroscopy (EELS) was used to investigate the volume plasmon energy losses in and around shear bands. Energy shifts of the peak centre and changes in the peak width (FWHM) reflecting the damping were precisely determined within an accuracy of a few meV using an open source python module (Hyperspy) to fit the shapes of the plasmon and zero-loss peaks with Lorentzian functions. The maximum bulk plasmon energy shifts were calculated for the bright and dark shear band segments relative to the matrix to be about 38 and 14 \,meV, respectively. The damping was observed to be larger for the denser regions. The analysis presented here suggests that the changes in the plasmons are caused by two contributions: (i) Variable damping in the shear band segments due to changes in the medium-range order (MRO). This affects the static structure factor $S(k)$, which, in turn, leads to either reduced or increased damping according to the Ziman-Baym formula. (ii) The ionic density and the effective electron mass appearing in the zero-momentum plasmon frequency formula $E_p(q=0)$ are coupled and give rise to small variations in the plasmon energy. The model predicts plasmon energy shifts in the order of meV. 
\end{abstract}


\end{frontmatter}


\section{Introduction}
Crystalline materials possess the ability to deform at constant volume along slip planes since the periodicity of the lattice provides identical atomic positions for the sheared material to lock in to. However, in the absence of a lattice, as for example in metallic glasses, this possibility does not exist. As a consequence, the mismatch between sheared zones (shear bands) and surrounding matrix needs to be accommodated by extra volume \cite{spaepen1977microscopic,Argon1979,donovan1981structure,heggen2005creation,klaumunzer2011probing,pan2011softening,greer2013shear,ma2015tuning}. Different experimental techniques have provided evidence that the extra volume is present in shear bands \cite{li2002nanometre, lechner2010vacancy, bokeloh2011tracer, bunz2014low, mitrofanov2014impact}. The sheared zones are thus softer than the surrounding matrix enabling the material to flow along them. Therefore, shear bands are associated with structural changes like local dilatation, implying a volume change and thus a change in density. An important issue is hence the local quantification of free volume inside shear bands. Recently, the local density within shear bands of different metallic glasses has been determined relative to the adjacent matrix using high angle dark field scanning transmission electron microscopy (HAADF-STEM) \cite{rosner2014density,schmidt2015quantitative,hieronymus2017shear,liu2017shear,hilke2019influence}. These experiments showed an alternation of higher and lower density regions along the propagation direction of the shear bands although on average shear bands were less dense than the surrounding matrix. The density changes along shear bands in Al$_{88}$Y$_{7}$Fe$_{5}$ correlated with small deflections along the propagation direction, compositional changes and structural changes in the medium-range order (MRO) \cite{rosner2014density,schmidt2015quantitative,hieronymus2017shear,liu2017shear,hilke2019influence,balachandran2019elemental,liu2019shear}. The observation of shear band regions which were denser than the surrounding matrix was initially somewhat unexpected since macroscopic measurements reported dilation only \cite{klaumunzer2011probing, lechner2010vacancy, shao2013high}. 

In this paper we focus on the changes in plasmon energy losses in a shear band of cold-rolled Al$_{88}$Y$_{7}$Fe$_{5}$ metallic glass. The plasmon energy shifts $\Delta E_p$ for both higher and lower density shear band segments were calculated relative to the surrounding matrix and found to be about 38 and 14 meV, respectively. The bulk plasmon energy shifts in the shear band are discussed on the basis of inelastic electron-phonon scattering. According to the Ziman-Baym theory, the different local MRO in the shear band segments will affect the first peak of the static structure factor $S(k)$ differently and hence affect the damping of the plasmon excitation differently. Moreover, the ionic density and effective electron mass appearing in the plasmon frequency formula at zero-momentum $E_p(q=0)$ are coupled. This gives rise to small variations in the plasmon energy $E_p(q=0)$ between the shear band segments and the matrix. 



\section{Experimental}
$\textrm{Al}_{88}\textrm{Y}_{7}\textrm{Fe}_{5}$ metallic glass is a marginal glass former. Melt-spun ribbons were produced by rapid quenching from the melt. Details can be found in reference \cite{bokeloh2010primary}. The amorphous state of the material was confirmed by x-ray diffraction (XRD) and selected area electron diffraction (SAED) prior to deformation. The ribbon material was deformed by cold-rolling yielding a thickness reduction of about $23 \, \%$. TEM specimens were prepared by twinjet electro-polishing using HNO$_3$:CH$_3$OH in a ratio 1:2 at $-22 \,^\circ$C applying voltages of about $-10.5 \,V$. Microstructural characterization was performed in an FEI S/TEM (Themis$^3$\,300) equipped with a high-brightness field-emission gun (X-FEG), quadruple energy-dispersive x-ray detectors (Super-X), HAADF detector (Fischione model 3000) and post-column electron energy loss spectrometer (Gatan Quantum 966 ERS imaging filter (GIF)) and operated at $200 \,$kV with an energy resolution of $0.8 \,$eV. Drift-corrected spectrum imaging \cite{hunt1991electron} was performed acquiring HAADF-STEM and EELS signals over an area of $50 \,$nm $ \times 200 \,$nm using a pixel size of $1 \,$nm $\times 1 \,$nm , a probe current of $60 \, $pA, a dwell-time of $3 \, $ms and a dispersion of $0.1 \, $eV. The $\alpha$- and $\beta$- semi-angles were $9.5$ and $6.9 \, $mrad, respectively. For the EDX elemental mapping a beam current of 4\,nA with a total acquisition time of $356\,$s was used. Plasmon signatures of sheared regions of Al$_{88}$Y$_{7}$Fe$_{5}$ metallic glass were analyzed using automated routines based on an open source python module (Hyperspy) \cite{de2019hyperspy} to fit the peak shapes of the zero-loss peak (ZLP) and the plasmon peak with Lorentzian functions \cite{mccomb1990characterisation} in order to determine their centre and width (FWHM) for each data point (pixel) of the spectrum image. The complete code and its description can be found in reference \cite{github2020hub}.


\section{Results}
\begin{figure}[htbp]
	\centering
	\includegraphics[width=0.8\columnwidth]{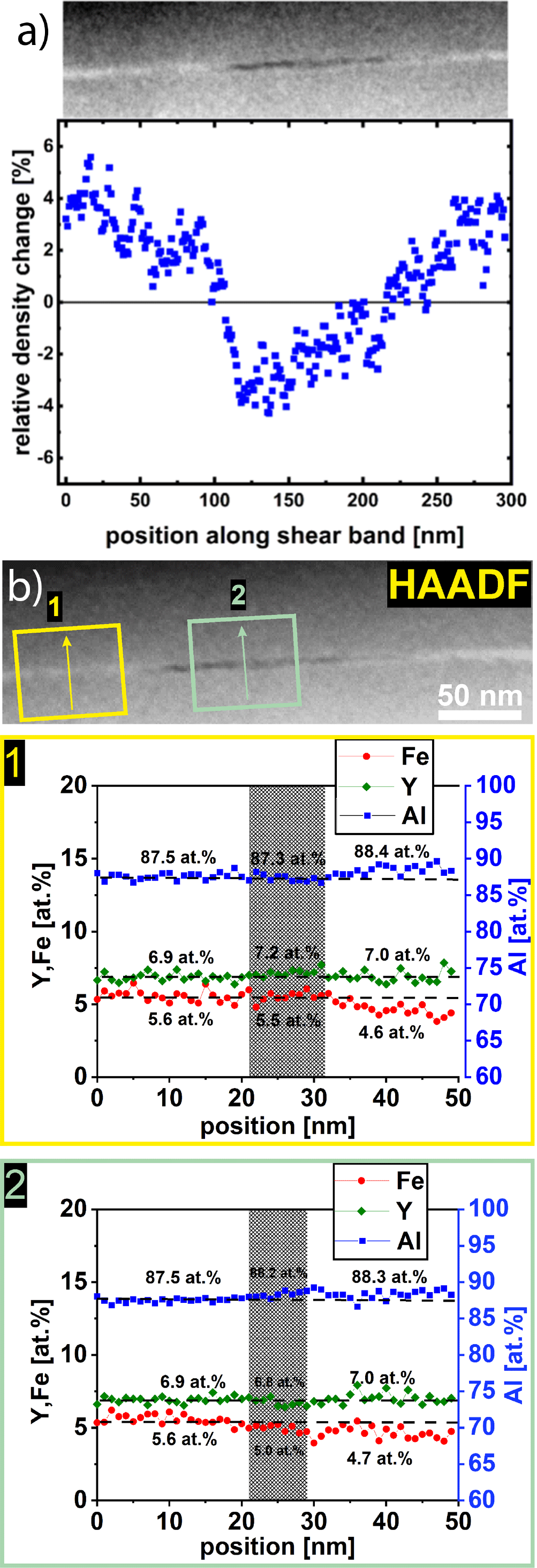}
	\caption{(a) Top: HAADF-STEM image of cold-rolled Al$_{88}$Y$_{7}$Fe$_{5}$ metallic glass showing a shear band with contrast reversal (bright-dark-bright). Bottom: Quantified density along the shear band. (b) Top: HAADF-STEM image showing the regions used for the EDX quantification. Below: Elemental profiles based on the EDX measurements extracted from the indicated boxes over the bright (1) and dark (2) shear band segment. The grey shaded box indicates the position of the shear band.}
	\label{FIG:1}
\end{figure}

In this study a representative part of a shear band (SB) in Al$_{88}$Y$_{7}$Fe$_{5}$ metallic glass was investigated by analytical TEM. The composition was analyzed by EDX using the newest generation of quadrupole detectors providing improved statistics for quantification \cite{schlossmacher2010nanoscale}. Fig.~\ref{FIG:1}a depicts a SB that exhibits a contrast change from bright to dark to bright. Slight deflections between the SB segments are correlated with the contrast variations \cite{rosner2014density, schmidt2015quantitative}. 

The contrast changes were quantified as density changes using the intensities of the HAADF-STEM signals as described in detail in references \cite{rosner2014density, schmidt2015quantitative, hieronymus2017shear, hilke2019influence}. The average density change for the bright SB segments is about $+3.9 \, \%$ and $-2.2 \, \%$ for the dark SB segment. The positive sign means densification and the negative sign dilation relative to the adjacent amorphous matrix. The regions used for the EDX chemical analysis are indicated by boxes in the HAADF image of Fig.~\ref{FIG:1}b.  In order to enhance the statistics of the elemental profiles the data was summed along the lines in the spectrum image parallel to the shear bands. The results of the compositional analysis for the different regions are shown in Fig.~\ref{FIG:1}b and are also listed in Tab.~\ref{TAB:1}. Both elemental profiles show a slight increase in Al and a decrease in Fe in the matrix on the right-hand side relative to the matrix on the left-hand side. Moreover, there is also a small compositional difference between the two SB segments. The bright SB segment is slightly depleted in Al and enriched in Fe and Y compared to the dark SB segment. However, the average atomic number Z does not vary largely ($\approx 1 \, \%$). Former investigations using fluctuation electron microscopy (FEM), a microscopical technique that is sensitive to MRO, revealed different degrees of nanoscale order. High structural order in terms of MRO was observed for the dark SB segments in $\textrm{Al}_{88}\textrm{Y}_{7}\textrm{Fe}_{5}$ while the bright SB segments showed even less structural ordering than the surrounding matrix. \cite{rosner2014density, schmidt2015quantitative}.  

The mean free path (MFP) was calculated for the nominal composition to be ($130.9 \pm 0.4$)\,nm according to Malis et al. \cite{malis1988eels} and ($162.3 \pm 2.1$)\,nm using the approach from Iakoubovskii et al. \cite{iakoubovskii2008thickness}. The values for the different local environments are summarized in Tab.~\ref{TAB:1}. 

\begin{figure}[htbp]
	\centering
	\includegraphics[width=\columnwidth]{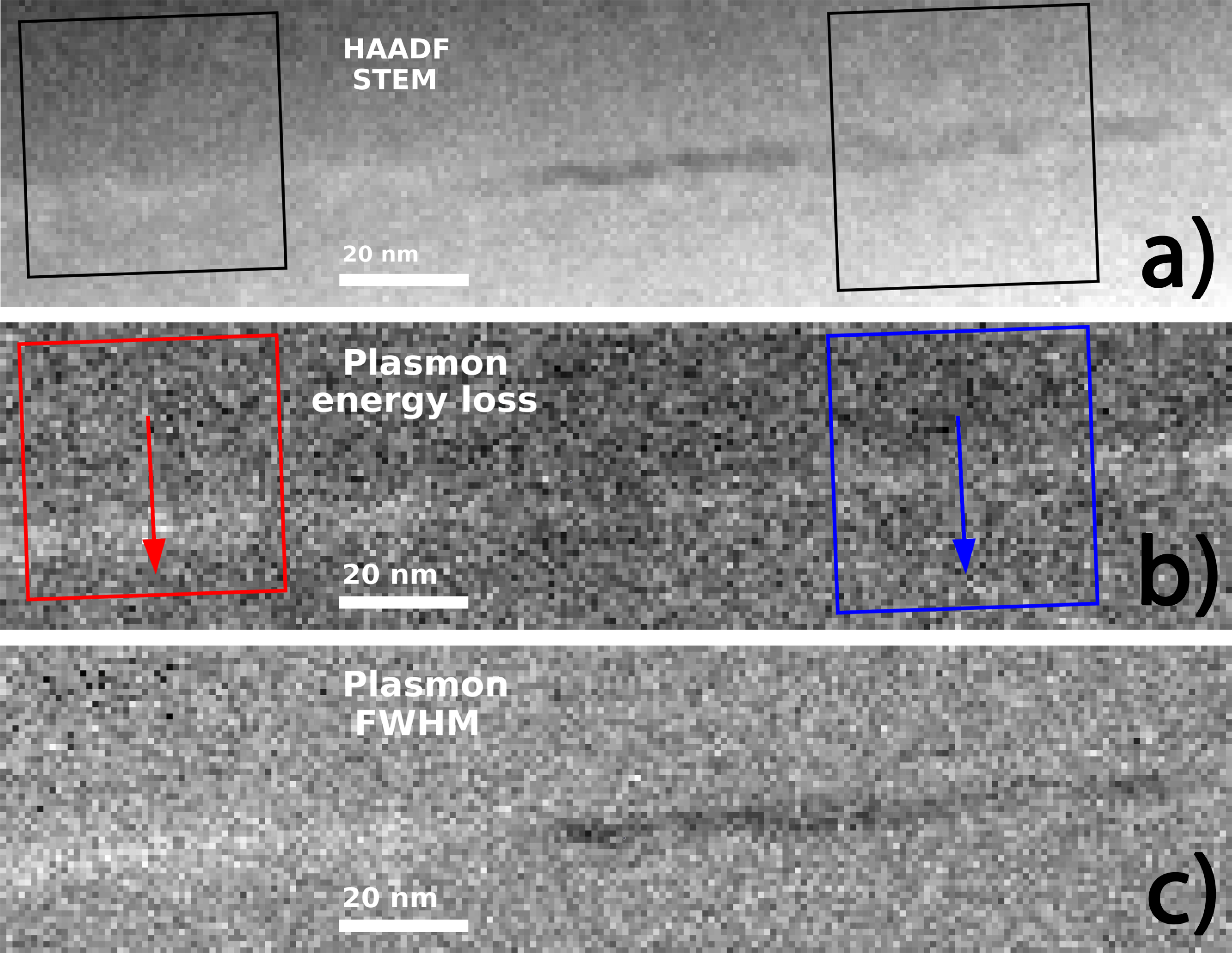}
	\caption{(a): HADDF-STEM signal from the spectrum image. (b): Map of the plasmon peak centre $E_{\textrm{max}}$. (c): Map of the plasmon peak width (FWHM) reflecting the damping.}
	\label{FIG:2}
\end{figure}

\noindent The volume plasmons in and around the SB were investigated by electron-energy loss (EEL) spectrum imaging. Fig.~\ref{FIG:2}a displays the HAADF signal from the spectrum image showing the location of the SB. From the Lorentzian functions fitted to the plasmon and zero-loss peaks of the individual EEL spectra, the peak centres E$_{max}$, and widths (FWHM), $\hbar\Gamma_p$, were determined. The map of the plasmon peak energy E$_{max}$ (Fig.~\ref{FIG:2}b) shows a noticeable energy shift relative to the matrix for both SB segments. The map of the plasmon peak width is shown in Fig.~\ref{FIG:2}c. Relative to the matrix, the plasmon peak is broader for the bright (denser) SB segment and narrower for the dark (less dense) SB segment.

\begin{figure}[h]
	\centering
	\includegraphics[width=\columnwidth]{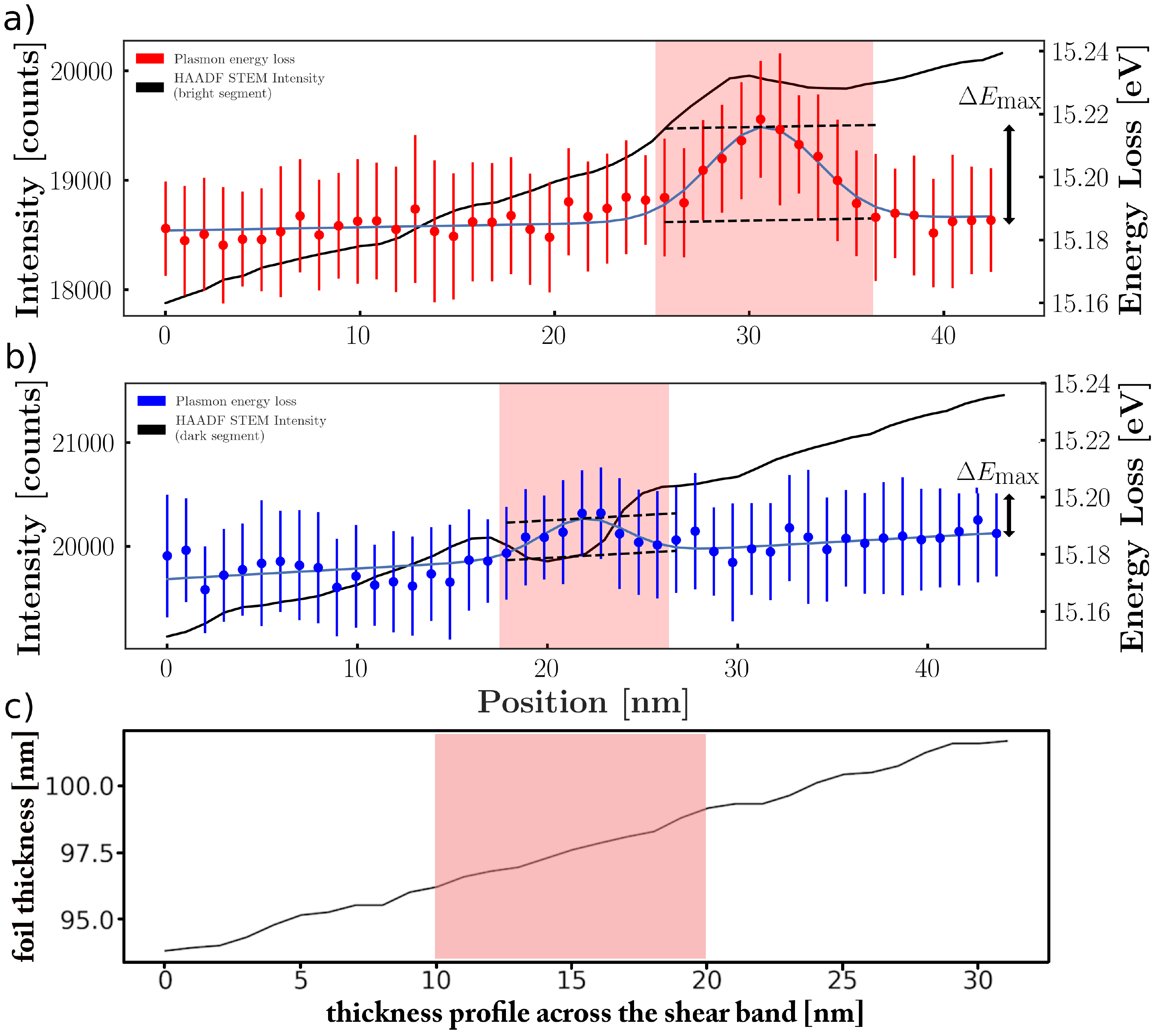}
	\caption{(a, b) Profiles (HAADF-STEM intensity and plasmon energy loss E$_{max}$) corresponding to the boxes shown in Fig.~\ref{FIG:2}b. The red and blue points correspond to the profiles across the bright (denser)and dark (less dense) SB segments. (c) Foil thickness profile drawn (top to bottom) across the SB shown in Fig.~\ref{FIG:2}a, respectively.}
	\label{FIG:3}
\end{figure}

The results of the quantification extracted from the boxed regions in Fig.~\ref{FIG:2}b are displayed in Fig.~\ref{FIG:3}a and Fig.~\ref{FIG:3}b together with the profile of the HAADF signal showing the SB position. Although the plasmon peak energy E$_{max}$ is somewhat variable in the matrix, it clearly shifts in the SB segments reaching a maximum value for E$_{max}$ of ($15.218 \pm 0.02$)\,eV for the bright SB segment and ($15.194 \pm 0.017$)\,eV for the dark SB segment. A summary of the results obtained from peak fitting are tabulated in Tab.~\ref{TAB:2}.  

A foil thickness profile drawn (top to bottom) across the investigated SB is displayed in Fig.~\ref{FIG:3}c. The slope of the thickness profile corresponds approximately to the slope observed in the HAADF signal (black line in Fig.~\ref{FIG:3}a and Fig.~\ref{FIG:3}b). Thus, it confirms the absence of preferential etching at the SB during sample preparation (electro-polishing). 

\begin{table*}[htbp]
	\centering
	\caption{Results of the EDX analyses for the two SB segments and matrix positions shown in Fig.~\ref{FIG:1}b. The atomic number, molar mass and mean free path (MFP) are calculated accordingly.}
	\begin{tabular}{l|ccc}
		Profile
		1 in Fig.~\ref{FIG:1}b                                                                                                                                                          & matrix
		(left) & bright
		segment & matrix
			(right)  \\ 
		\hline
		Al				[at.\%]                                                                                                                                   & 87.5              & 87.3               & 88.4                                   \\
		Fe				[at.\%]                                                                                                                                                                     & 5.6               & 5.5                & 4.6                                   \\
		Y [at.\%]                                                                                                                                       & 6.9               & 7.2                & 7.0                                    \\ \hline
		Average
		atomic number  Z                                                                                                                                                      & 15.52             & 15.59              & 15.42                                  \\ \hline
		Molar				mass [g/mol]                                                                                                                           & 32.87             & 33.03              & 32.64                                  \\ \hline
		\begin{tabular}[c]{@{}l@{}}				Mean				free path [nm] \\calculated after Malis [30]			\end{tabular}                                                                               & 130.8             & 130.7              & 131.1                                  \\ \hline
		\begin{tabular}[c]{@{}l@{}}				Mean				free path [nm] \\calculated after Iakoubovskii [31]			\end{tabular} & 162.1             & 160.8              & 162.4                                \\ \hline \hline
	\end{tabular}
	\centering
	\begin{tabular}{l|ccc}
		Profile
		2 in Fig.~\ref{FIG:1}b                                                                                                                                                          & matrix
		(left) & dark
		segment & matrix
			(right)  \\ 
		\hline
		Al				[at.\%]                                                                                                                                   & 87.5              & 88.2             & 88.3                                   \\
		Fe				[at.\%]                                                                                                                                                                     & 5.6               & 5.0              & 4.7                                    \\
		Y [at.\%]                                                                                                                                       & 6.9               & 6.8              & 7.0                                    \\ \hline
		average
		atomic number Z                                                                                                                                                       & 15.52             & 15.42            & 15.42                                  \\ \hline
		Molar				mass [g/mol]                                                                                                                           & 32.87             & 32.64            & 32.64                                  \\ \hline
		\begin{tabular}[c]{@{}l@{}}				Mean				free path [nm] \\calculated				after Malis [30]			\end{tabular}                                                                            & 130.8             & 131.1            & 131                                    \\ \hline
		\begin{tabular}[c]{@{}l@{}}				Mean				free path [nm] \\calculated after Iakoubovskii [31]			\end{tabular} & 162.1             & 163.8            & 162.4 \\ \hline \hline                                
	\end{tabular}
	\label{TAB:1}
\end{table*}

\begin{table*}[htbp]
	\centering
	\caption{Calculated maximum and standard deviation based on the Lorentzian peak fitting for the plasmon energy loss (peak centre) $E_\textrm{max}$ and peak width $\hbar\Gamma_p$ at FWHM. The expected bulk plasmon energy loss $E_p (q=0)$ for undamped plasmons calculated using Eq.~\ref{EQ6} and the difference between SB segments and the adjacent matrix are also shown.}
	\begin{tabular}{l||c|c|c|c}
		& $E_{\textrm{max}}$ {[}eV]            & $\hbar\Gamma_p${[}eV]           & $E_p(q=0)${[}eV]            & $\Delta E_p$ {[}meV]        \\ \hline \hline
		bright
		SB segment                                                                   & $(15.218 \pm 0.020)$ & $(2.796 \pm 0.062)$ & $(15.282 \pm 0.02)$ & $(37.7 \pm 20.5)$  \\
		\begin{tabular}[c]{@{}l@{}}			Adjacent			matrix (averaged) \\bright SB segment 		\end{tabular}     & $(15.185 \pm 0.004)$ & $(2.688 \pm 0.017)$ & $(15.244 \pm 0.004)$ &         -       \\
		dark SB segment                                                                     & $(15.194 \pm 0.017)$ & $(2.63 \pm 0.021)$ & $(15.251 \pm 0.018)$ & $(13.8 \pm 18.5)$   \\
		\begin{tabular}[c]{@{}l@{}}			Adjacent			matrix (averaged) \\dark			SB segment 					\end{tabular} & $(15.179 \pm 0.006)$ & $(2.679 \pm 0.009)$ & $(15.238 \pm 0.006)$ &          -     
	\end{tabular}
	\label{TAB:2}
\end{table*}

\begin{table*}[htbp]
	\centering
	\caption{Recalculated values using processed data by applying the Fourier-log deconvolution method to the low-loss spectra in comparison to the unprocessed data shown in Tab.~\ref{TAB:2}.}
	\begin{tabular}{l||c|c|c|c}
		& $E_{\textrm{max}}$ {[}eV]            & $\hbar\Gamma_p${[}eV]           & $E_p(q=0)${[}eV]            & $\Delta E_p$ {[}meV]        \\ \hline \hline
		bright
		SB segment                                                                   & $(15.247 \pm 0.020)$ & $(2.833 \pm 0.036)$ & $(15.313 \pm 0.02)$ & $(33.2 \pm 20.6)$  \\
		\begin{tabular}[c]{@{}l@{}}			Adjacent			matrix (averaged) \\bright SB segment 		\end{tabular}     & $(15.218 \pm 0.005)$ & $(2.737 \pm 0.016)$ & $(15.279 \pm 0.005)$ &         -       \\
		dark SB segment                                                                     & $(15.215 \pm 0.019)$ & $(2.648 \pm 0.033)$ & $(15.272 \pm 0.019)$ & $(10.9 \pm 19.4)$   \\
		\begin{tabular}[c]{@{}l@{}}			Adjacent			matrix (averaged) \\dark			SB segment 					\end{tabular} & $(15.202 \pm 0.004)$ & $(2.694 \pm 0.017)$ & $(15.262 \pm 0.004)$ &          -     
	\end{tabular}
	\label{TAB:3}
\end{table*}

\section{Discussion}
In the following section the experimental results are discussed.

\subsection{Effect of the background and plural scattering on the plasmon energy shift}
\begin{figure}[htbp]
	\centering
	\includegraphics[width=\columnwidth]{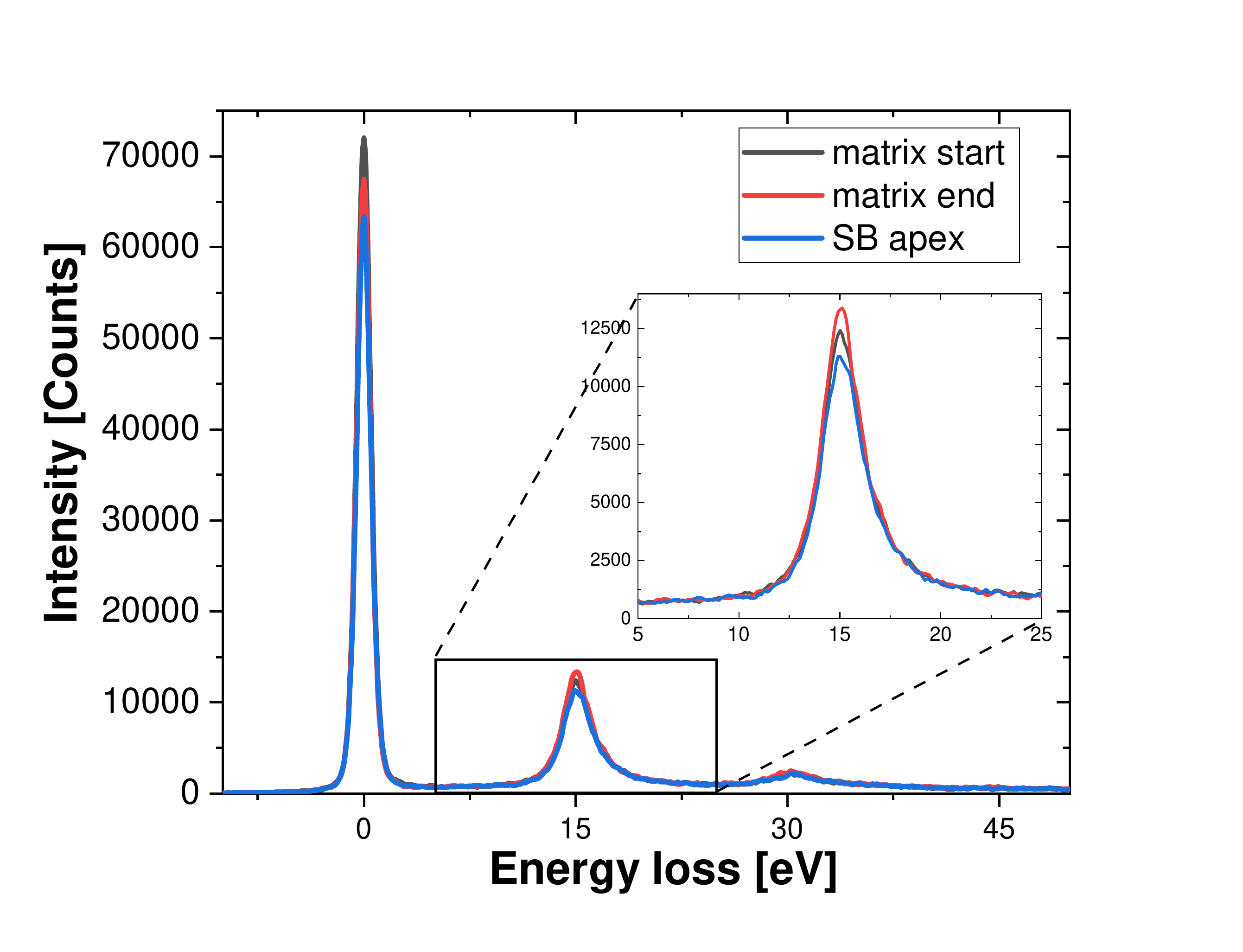}
	\caption{Three representative EEL spectra (raw data) taken from the linescan shown in Fig.~\ref{FIG:3}a representing the start (matrix black), the end (matrix red) and the apex of bright SB segment (blue).}
	\label{FIG:4}
\end{figure}

To discuss the influence of the background and the degree of plural scattering on the energy shifts and the peak broadening, three representative spectra (raw data) are depicted in Fig.~\ref{FIG:4} which are taken from different positions of the linescan shown in Fig.~\ref{FIG:3}a. A small second plasmon peak can be noticed in all spectra indicating the presence of plural scattering due to the foil thickness, which ranges from 92 - 102 nm (see Fig.~\ref{FIG:3}c) or 0.7 - 0.8 mean free path. While a visual inspection of Fig.~\ref{FIG:4} shows no apparent change in the background, an indiscernible effect may arise from subtle changes in the second plasmon peak. To evaluate this contribution, plural scattering was removed using the Fourier-log deconvolution technique \cite{johnson1974determination, egerton1985fourier, egerton1988use}. Subsequently, all values were recalculated using deconvoluted spectra and the results are given in Tab.~\ref{TAB:3}. A comparison of the data in Tabs.~\ref{TAB:2} and~\ref{TAB:3} shows that, while there are small changes in the values as a result of deconvolution, the trends for the peak shifts and widths remain the same. From this we conclude that in our case the influence of background and plural scattering are minor relative to the total error and thus justifies the use of unprocessed low-loss data, which are preferred for practical reasons.

\subsection{Compositional contributions}
\noindent While there are small changes in the composition of the matrix and the SB regions (Fig.~\ref{FIG:1}b), there is no apparent correlation with the shifts in the plasmon peak energy (Figs.~\ref{FIG:3}a and b). According to the experimental work of Hibbert et al. \cite{hibbert1972variation}, the variation of the plasmon energy loss with composition in Al-rich Al-Mg solid solutions showed that an addition of 1 at.\% Mg as impurity caused a decrease in the plasmon energy loss of about -50 meV.  Considering the additions of Fe and Y in the first instance to have similar effects to the plasmon energy loss of Al as Mg, a compositional difference of 1 at.\% Al between the bright and dark SB segments (see Tab.~\ref{TAB:1}) should then result in a similar energy shift of about - 50 meV. However, in our case the bright SB segment in Fig.~\ref{FIG:1}b, which has less Al, shows an increased energy shift of about 24 meV. Thus, the compositional changes are insufficient to explain the observed plasmon energy shift in the SB segments.

\subsection{Theoretical analysis of plasmon resonance enhancement in shear bands}
The plasmon resonance peak may be modeled as a Lorentzian function centered at the plasmon energy loss $E_{\textrm{max}} \approx 15.2 \,$eV and with FWHM in the order of $\hbar\Gamma_p \approx 2.7 \, $eV (see Tab.~\ref{TAB:2}), where $\Gamma_p$ denotes the damping coefficient of the plasmon excitation. The observed values are close to typical values for crystalline Al, i.e. $E_p = 15\,$ eV \cite{egerton2011electron}. However, it is worth noting that crystalline Al shows a peak shift of about $0.4 \, $eV upon melting into a liquid \cite{faber2010introduction}.

 By definition, the electrons that contribute to the volume plasmon peak are free or nearly-free electrons. We underline the fact that in our material, which is to a great extent Al, the nearly-free electron model is expected to work fairly well, unlike the case in alloys where transition metals are dominant. Taking the formula for the bulk plasmon energy loss, $E_p=\hbar \sqrt{\frac{n e^2}{m \epsilon_0}}$, where $\hbar$ is the reduced Planck constant, $n$ is the electron density, $e$ and $m$ are the electron charge and mass, and $\epsilon_0$ is the vacuum permittivity, the shifts of the plasmon energy losses in the SB cannot be explained simply in terms of a uniform nearly-free electron concentration over the entire sample. Thus, although the electrons are treated as delocalized in the nearly-free electron model, there are differences in concentration of the nearly-free electrons in different parts of the sample. These differences reflect the underlying differences of ionic density and the individual damping characteristics of the different regions. Thus, the measured plasmon energy loss in our material is not given by $E_{p} = \hbar \sqrt{\frac{n e^2}{m \epsilon_0}}$, because this simple estimate holds only in the absence of damping (hence, it works only for perfect defect-free crystalline metals at low temperature). In a metallic glass at room temperature, damping is of course a very important contribution to the maximum, $E_\textrm{max}$, occurring. According to theory \cite{nikjoo2012interaction, egerton2011electron}, the maximum occurs at an energy approximately proportional to the inverse of the imaginary part of the dielectric function, given by \cite{egerton2011electron}

\begin{equation*}
\textrm{max}\{ \Im\left(\frac{-1}{\epsilon(\omega)}\right) \} \approx \frac{\hbar \omega_p}{\Gamma_p} = \frac{E_p}{\Gamma_p}
\end{equation*}
and reaches a maximum value of $\frac{\omega_p}{\Gamma_p}$ at an energy loss 
\begin{equation}
E_{\textrm{max}} = \left[(E_p)^2 - \left(\frac{\hbar\Gamma_p}{2}\right)^2\right]^{\frac{1}{2}} 
\label{EQ1},
\end{equation}

\noindent which is close to $E_\textrm{p}$, in the case of small damping. In a real system, the measured plasmon energy loss is $E_\textrm{max}$, not $E_p$ and the two values coincide only for ideal perfect crystals at low temperature. The undamped plasmon frequency is related to the undamped plasmon energy loss via $\hbar \omega_p = E_p$. The damping coefficient $\Gamma_p$ depends on the local scattering events between conduction electrons and ion-cores and hence the local microstructure. Therefore, $\Gamma_p$ may vary spatially because in some portions of the material scattering events are more frequent than in other parts of the material (cf. Fig.~\ref{FIG:2}c). In general, there are several contributions to damping besides ion scattering: important effects are Landau damping, inter-band transitions \cite{gelin2016anomalous} and electron-phonon scattering. It should be mentioned that damping of collective excitations in amorphous solids may be strongly non-local: as recently reported, long-range (power-law) stress correlations in amorphous solids cause logarithmic decay of phonon damping with the wave vector \cite{march1984coulomb}. However, for Al-rich materials the Fermi energy lies well below the band top so that single-particle excitations are very unlikely, and damping can be assumed to be caused predominantly by electron-ion and electron-phonon scattering. Hence, the damping coefficient may be estimated from its definition in terms of the relaxation time (mean time between scattering collisions with ions), $\Gamma_p = \tau^{-1}$. The relaxation time $\tau$ is related to the resistivity $\rho^*$ via $\rho^* = \left( \frac{n e^2 \tau}{m} \right)^{-1}$.

The resistivity of metallic glasses is described by the Ziman theory \cite{faber2010introduction, march1984coulomb}, which was originally developed for liquid metals but works also for certain glasses. The famous Ziman formula for the resistivity of amorphous metals reads 
\begin{equation}
\rho^* = \frac{3 \pi m^2}{4 n_i e^2 \hbar^3 k_F^6} \int_{0}^{2 k_F} k^3 S(k) |V(k)|^2 dk 
\label{EQ2}
\end{equation}
which gives the damping coefficient as \cite{nikjoo2012interaction}:
\begin{equation}
\Gamma_p = \tau^{-1} = \frac{3 \pi n m}{4 n_i \hbar^3 k_F^6} \int_{0}^{2 k_F} k^3 S(k) |V(k)|^2 dk 
\label{EQ3}
\end{equation}

\noindent where $k$ denotes the wave vector (here defined as the reciprocal of the radial distance $r$ measured from one ion taken as the centre of a spherical frame and not to be confused with the wave vector of the incident electron beam), $k_F$ is the Fermi wave vector, and $n_i$ is the number density of ions in the material. Furthermore, $S(k)$ is the static structure factor of the metallic glass (i.e. the spatial Fourier transform of the radial distribution function $g(r)$), which again is a \textit{local} quantity that significantly differs for the bright and dark SB segments, since the local atomic structure (topological order) is different for the two segments \cite{schmidt2015quantitative}. Finally, $V(k)$ is the average Thomas-Fermi-screened electron-ion pseudopotential form factor for elastic scattering at the Fermi surface \cite{march1984coulomb}. The Ziman formula accounts for elastic electron-ion scattering only, and thus is typically valid at temperatures well below the Debye temperature. Since the Debye temperature for $\textrm{Al}_{88}\textrm{Y}_{7}\textrm{Fe}_{5}$ is $(360 \pm 6)\,$K \cite{Gerlitz2018}, the Ziman formula should be replaced by the Baym formula \cite{march1984coulomb}. Thus $S(k)$ in Eq.~\ref{EQ3} needs to be replaced by a frequency integral over the dynamic structure factor $S(k, \omega)$ times a factor $\frac{\hbar \omega}{kT (\exp{\frac{\hbar \omega}{kT}) - 1}}$. Using the Vineyard approximation $S(k, \omega) S(k) S_s(k, \omega)$, where $S_s(k, \omega)$ denotes the self-part of the atomic dynamics, the dynamic structure factor $S(k, \omega)$ can be extracted from the frequency integral, and upon performing the frequency integral, the damping coefficient becomes 
\begin{equation}
\Gamma_p = \tau^{-1} = \frac{3 \pi n m}{4 n_i \hbar^3 k_F^6} \int_{0}^{2 k_F} k^3 S(k) g(k) |V(k)|^2 dk 
\label{EQ4}
\end{equation}
where $g(k)$ is a function of $k$ only. As a first approximation, the integral is dominated by $S(k)$, especially for the low-$k$ region. Alternatively, one can manipulate the Baym formula after Meisel and Cote \cite{meisel1977structure} for metallic glass and arrive at the same simplification.

Regarding the integration limit above, it is important to note that for some metallic glasses the Nagel-Tauc rule \cite{nagel1975nearly} for metallic glass stability and formability stipulates that $2 k_F \approx k_\textrm{max}$, where $k_\textrm{max}$ denotes the wave vector of the first peak in the structure factor. However, for pure Al it is known that $2 k_f \approx k_\textrm{min}$, where $k_\textrm{min}$ is the minimum after the first peak in $S(k)$, as is the case for all three-valent metals \cite{march1984coulomb}. Since our system is very rich in Al, it is likely that the value of $k$, which satisfies $2 k_F \approx k$, lies somewhere between  $k_\textrm{min}$ and $k_\textrm{max}$. 

\begin{figure}[h]
	\centering
	\includegraphics[width=\columnwidth]{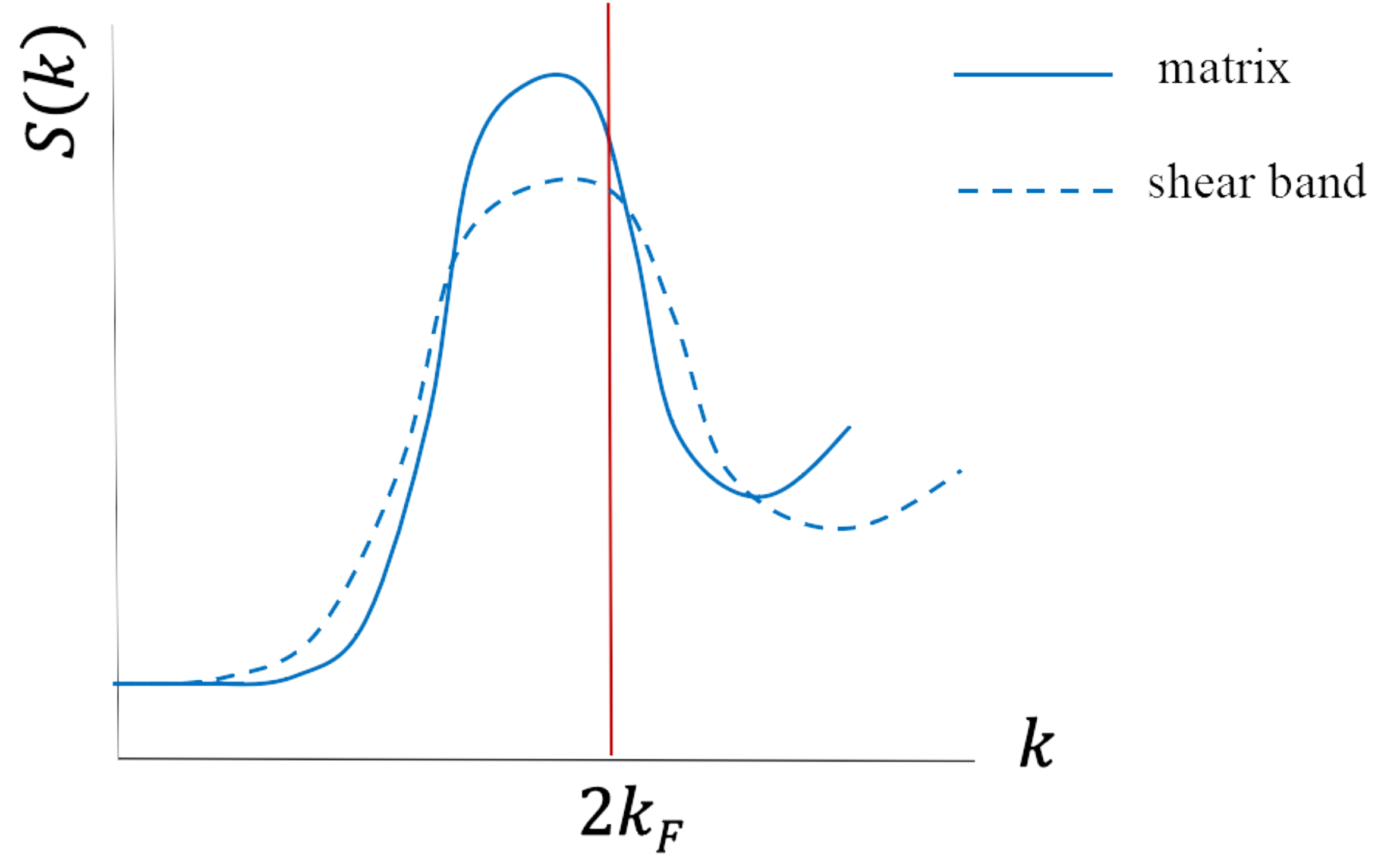}
	\caption{Schematic illustration of the static structure factor $S(k)$ for metallic glass. $S(k)$ is the spatial Fourier transform of the radial distribution function g(r). Pronounced structural order in terms of medium-range order within the dark shear band (dashed line) leads to broadening and lowering of the first peak of $S(k)$ relative to the matrix. The contribution to the integral in the Ziman-Baym formula Eq.~\ref{EQ4} leads to less plasmon damping in the dark SB.}
	\label{FIG:5}
\end{figure}

Since the pseudopotential $V(k)$ is approximately given by the Fourier transform of the Thomas-Fermi screened Coulomb attraction, the function $V(k)$ is relatively shallow within the range of integration. Hence, the integral is dominated by the first peak in the static structure factor $S(k)$, i.e. by the fraction of short and medium-range order in the atomic coordination (Fig.~\ref{FIG:5}). 

\subsection{Dependence of the damping coefficient on the local microstructure}
The dependence of the damping coefficient on the local free volume is clearly the same as the volume- (or equivalently, density-) dependence of the resistivity. The effect of dilation on resistivity has been studied extensively by Ziman and Faber on the basis of Eq.~\ref{EQ2}. In particular, Faber \cite{faber2010introduction} has provided the following expression for the volume-dependence of the resistivity:
\begin{equation}
\begin{split}
\frac{V}{\rho^*} \left( \frac{\partial \rho^*}{\partial V} \right)_T &= \frac{2}{3} \xi - 1\\
&+ \frac{\int_{0}^{2 k_F} k^3 \left( \frac{\partial S(k)}{\partial V} \right)_T |V(k)|^2 dk}{\int_{0}^{2 k_F} k^3 S(k) |V(k)|^2 dk} \\
&+ \frac{\int_{0}^{2 k_F} k^3 \left( \frac{\partial |V(k)|^2}{\partial V} \right)_T S(k) dk}{\int_{0}^{2 k_F} k^3 S(k) |V(k)|^2 dk} 
\end{split}
\label{EQ5}
\end{equation}
\noindent where $\xi = -\frac{k_F}{2 \rho} \left( \frac{\partial \rho}{\partial k_F} \right)$ is a dimensionless parameter. Eq.~\ref{EQ5} is attractive because it clearly singles out the three main contributions to the change of resistivity upon dilation: i) the effect of expanding the Fermi sphere, encoded in $\xi$; ii) the change in short and medium-range order encoded in the derivative $\frac{\partial S(k)}{\partial V}$; iii) the effect on the pseudopotential of any change in the screening properties of the conduction electron gas. Faber \cite{faber2010introduction} concluded from his analysis that the second term in Eq.~\ref{EQ5} is the one which usually dominates the overall dependence of resistivity on dilation. Assuming that the position and the width of the first peak of $S(k)$ scale with $V^{-1/3}$, Faber concluded that $\frac{\partial S(k)}{\partial V}\sim \frac{\partial S(k)}{\partial k}$.

In the absence of a more quantitative theory, we resort to the following argument based on the consideration of the microstructure in the SB. In the Ziman-Baym formula  $S(k)$ is related to the pair-correlation function or radial distribution function, RDF in isotropic systems, giving the averaged probability of finding any atom at a distance r from the atom at the centre of the frame. 

We now apply the Ziman-Baym theory to estimate the change in the plasmon energy loss in the SB with respect to the matrix due to damping. We know from the plasmon peak fitting of the EELS data that the FWHM in the matrix adjacent to the bright SB segment is $\hbar\Gamma^{matrix}_\textrm{bright} = 2.688 \,$eV, whereas in the matrix adjacent to the dark SB segment it is $\hbar\Gamma^{matrix}_\textrm{dark} = 2.679 \, $eV. For the SB segments we obtain $2.796 \, $eV and $2.63 \, $eV, for bright and dark, respectively. The damping of the excitation is reduced in the dark SB segment due to higher structural order in terms of MRO (small crystal-like Al-rich clusters embedded in amorphous material) resulting in a broader and lowered first peak of $S(k)$ \cite{rosner2014density, schmidt2015quantitative}, which in turn results in a reduced resistivity (hence a lower damping of the plasmon), because the lower peak in $S(k)$ gives a smaller value for the integral in Eq.~\ref{EQ4}. For the bright SB segment the damping is higher than in the matrix because there is even less ‘structure’ in terms of MRO and hence less dispersion with the result that the first peak of $S(k)$ is higher.

\subsection{Model estimate of plasmon energy shift in the shear bands }
Now, the energy of the plasmon loss is given by Eq.~\ref{EQ1} as
\begin{equation}
E_p(q=0)=\sqrt{E_{\textrm{max}}^2+\left(\frac{\hbar\Gamma_p}{2}\right)^2},
\label{EQ6}
\end{equation}
and for the relative shift in each segment we have then
\begin{equation}
\begin{split}
E_{p}^{\textrm{SB}} \left(q=0 \right)& - E_{p}^{\textrm{matrix}} \left(q=0 \right) \\ 
&= \sqrt{\left[\left(E_{\textrm{max}}^{\textrm{SB}}\right)^2 + \left(\frac{\hbar\Gamma_p}{2}\right)^2 \right]} \\
&-  \sqrt{\left[\left(E_{\textrm{max}}^{\textrm{matrix}}\right)^2 + \left(\frac{\hbar\Gamma_p}{2} \right)^2 \right]},
\end{split}
\label{EQ7}
\end{equation}
where $E_p$ denotes the bulk plasmon energy loss in the complete absence of damping or at momentum transfer $q=0$. 

Using the measured values of $\hbar\Gamma_p$ in the various regions together with the measured values of the plasmon energy loss $E_{\textrm{max}}$, one can infer the values of undamped plasmon energy $E_{p,0}^{\textrm{SB}}$ in the SB segments and compare those values to the value measured in the matrix $E_{p,0}^{\textrm{matrix}}$. These values are then related to the ionic density  and the effective electron mass $m^{*}$ in the various regions via $E_{p,0} = \hbar \sqrt{\frac{n_e \cdot e^2}{m^{*} \cdot \epsilon_0}} = \hbar \sqrt{\frac{n_i \cdot \left(ze^2\right)}{m^{*} \cdot \epsilon_0}}$. Neglecting the variation of electron effective mass in the first instance, we arrive at the following enhancement for the bright (densified) SB segment: 
$\frac{E_{p,0}^{\textrm{SB,bright}}}{E_{p,0}^{\textrm{matrix}}} = 1.0014 \approx \sqrt{\frac{n_{i}^{\textrm{SB,bright}}}{n_{i}^{\textrm{matrix}}}}$, and for the dark (dilated) SB segment we get: $\frac{E_{p,0}^{\textrm{SB,dark}}}{E_{p,0}^{\textrm{matrix}}}  = 1.0005 \approx \sqrt{\frac{n_{i}^{\textrm{SB,dark}}}{n_{i}^{\textrm{matrix}}}}$. 

\noindent This result can be explained as follows. The electron effective mass in metallic glasses is controlled by electron-phonon coupling via \cite{faber2010introduction}: $m^{*} = m\left(1+\lambda\right)$, where the mass-enhancement factor (electron-phonon coupling parameter) is given through the Eliashberg function, valid also for metallic glasses \cite{setty2020effective}, $\lambda = \int_{0}^{\omega_{\textrm{D}}} {\frac{d\omega}{\omega}}  \alpha^{2} F(\omega)$. Here, $\alpha$ is the electron-phonon matrix element and $F(\omega)$ is the vibrational density of states (VDOS). The above formula for the mass-enhancement factor $\lambda$ can be expressed in terms of an integral over the dynamical structure factor $S(k, \omega)$ times the matrix element via the Eliashberg theory. Meisel and Cote \cite{meisel1981eliashberg} and independently Jaeckle and Froboese \cite{jackle1980electron} using a slightly different derivation, have shown that in metallic glasses $\lambda \propto \Lambda^{-1}$, where $\Lambda$ denotes the mean-free path of the electron in the ionic environment. The prefactors in this relation are density-independent constants, hence they do not change from region to region, whereas the mean free path is roughly inversely proportional to the local ionic density $n_i^{\textrm{SB}}$ in the region. The mean free path was calculated for the different regions according to Malis et al. \cite{malis1988eels} and found to be around ($130.9\pm 0.4$) nm. Using the model (Kramers-Kronig sum rule) of Iakoubovskii et al. \cite{iakoubovskii2008thickness} we find a mean free path of ($162.3\pm 2.1$) nm. While the absolute values of the mean free path differ to a great deal depending on the used model \cite{malis1988eels, iakoubovskii2008thickness}, the values calculated within one model do not vary significantly in space (Tab.~\ref{TAB:1}). Thus, assuming the mean free path to be constant, we get 
\begin{equation}
\frac{E_{p,0}^{\textrm{SB}}}{E_{p,0}^{\textrm{matrix}}} \approx \sqrt{\frac{n_{i}^{\textrm{SB}}}{n_{i}^{\textrm{matrix}}}} \sqrt{\frac{1+ {\lambda}^{\textrm{matrix}}}{1+{\lambda}^{\textrm{SB}}}}.
\label{EQ8}
\end{equation}
Moreover, since in metallic glasses the mass-enhancement parameter is large ($\lambda \gg 1$), it is clear that $\frac{\lambda^{\textrm{matrix}}}{\lambda^{\textrm{SB}}} \propto \frac{\Lambda^{\textrm{SB}}}{\Lambda^{\textrm{matrix}}} \propto \frac{n_{i}^{\textrm{matrix}}}{n_{i}^{\textrm{SB}}}$ in the second radical of Eq.~\ref{EQ8} almost cancels the effect of the first radical. 
Accordingly, the plasmon energy shift for the dark SB segment is mostly related to a reduced damping. In fact, the increased MRO in the dark SB segment gives rise to a reduced first peak of $S(k)$ and thus leads to reduced damping because of the smaller contribution of the lowered peak of $S(k)$ to the integral in Eq.~\ref{EQ2}. Somewhat different is the case of the bright SB segment, where the coupling of effects is comparatively more effective and leaves a slightly larger enhancement for the value of $E_{p,0}^{\textrm{SB}}$ with respect to the matrix due to the ionic density.

In essence, we have two contributions that are responsible for the plasmon energy shifts; that is, (i) damping due to electron-phonon scattering and (ii) the ionic density. The increase in damping for the denser (bright) SB segment and the decrease in the dilated (dark) one with respect to the matrix is related to the level of structural order present in the different regions \cite{rosner2014density, schmidt2015quantitative, hilke2019influence, hilke2020role, Davani2019}. This also fits to recent findings observed in granular media where sound damping is determined by the interplay between elastic heterogeneities and inelastic interactions \cite{saitoh2020sound}. The second contribution is related to the efficacy of the coupling between the ionic density and the effective electron mass appearing in the plasmon frequency formula at zero-momentum $E_p(q=0)$, which is less effective in the dark SB segment than in the bright one.

\section{Conclusions}
Shifts and widths of plasmon energy losses were experimentally determined from a sheared zone (shear band and its immediate environment) of a metallic glass using automated routines based on an open source python module (Hyperspy) to fit the peak shapes of the zero-loss peak and the plasmon peaks \cite{github2020hub}. These signals are characteristic fingerprints and therefore suitable to visualize deformation features in amorphous materials such as shear bands. The model presented here suggests two reasons for the the plasmon energy shifts. First, variable plasmon damping in the shear band segments caused by differences in the medium-range order present. This affects the first peak of the static structure factor $S(k)$, which leads to either lowered or increased damping according to the Ziman-Baym resistivity formula. The second reason is that the ionic density and the effective electron mass appearing in the zero-momentum plasmon frequency formula $E_p(q=0)$ are coupled and give rise to small variations in the plasmon energy between the shear band and the matrix. The model predicts plasmon energy shifts in the order of meV \cite{huang2020extracting}.

\section{Acknowledgments}
We gratefully acknowledge financial support from the Deutsche Forschungsgemeinschaft (WI 1899/29-1; project number 325408982). We thank Dr. P. Schlossmacher (Thermo Fisher Scientific, FEI Deutschland GmbH) for enabling the measurements at the Nanoport in Eindhoven during a demonstration of a Themis$^3$\,300 microscope. 
A.Z. gratefully acknowledges financial support from the US Army Research Office through contract no. W911NF-19-2-0055. Fruitful discussions with Drs. Vitalij Hieronymus-Schmidt and Sven Hilke are acknowledged.

\bibliography{LITERATURE-Manuscript-Plasmon-AlYFe_Ultra_AZ}

\end{document}